\documentclass[12pt]{iopart}

\usepackage{graphicx}

\begin{document}

\title{Time-dependent transport in Aharonov-Bohm interferometers}

\author{V. Kotim\"aki$^1$, E. Cicek$^2$,
A. Siddiki$^{3,4}$ and E. R{\"a}s{\"a}nen$^{1,4}$}

\address{$^1$ Nanoscience Center, Department of Physics, University of Jyv{\"a}skyl{\"a}, FI-40014 Jyv{\"a}skyl{\"a}, Finland}
\address{$^2$ Trakya University, Department of Physics, 22030 Edirne, Turkey}
\address{$^3$ Physics Department, Faculty of Science, Istanbul University, 34134 Vezneciler-Istanbul, Turkey}
\address{$^4$ Physics Department, Harvard University, 02138 Cambridge MA, USA}

\begin{abstract}
A numerical approach is employed to explain
transport characteristics in realistic, quantum Hall based
Aharonov-Bohm interferometers.
First, the spatial distribution of incompressible strips, and thus
the current channels, are obtained applying a self-consistent Thomas-Fermi
method to a realistic heterostructure under quantized Hall conditions.
Second, the time-dependent Schr\"odinger equation is solved for
electrons injected in the current channels. Distinctive Aharonov-Bohm
oscillations are found as a function of the magnetic flux.
The oscillation amplitude strongly depends on the mutual distance between the
transport channels and on their width. At an optimal distance
the amplitude and thus the interchannel transport is maximized, which
determines the maximum visibility condition.
On the other hand, the transport is fully suppressed at magnetic
fields corresponding to half-integer flux quanta. The
results confirm the applicability of realistic Aharonov-Bohm
interferometers as controllable current switches.
\end{abstract}


\maketitle

\section{Introduction}

The Aharonov-Bohm (AB) effect~\cite{Aharonov:59} is among the most
significant and useful phenomena in quantum mechanics.
The AB effect manifests itself in the interaction between a charged
particle and an electromagnetic field, even if the {\em local}
magnetic and electric fields are zero in that region.
The necessary information is included in the vector
potential $\mathbf{A}$, which induces a phase shift
in the wave function of the electron traveling
along a specific path. In a double-slit system,
or in a quantum ring (see Ref.~\cite{Esa:ab} and references therein),
the relative phase shift between two electrons traveling along different paths is
$\Delta \phi = 2\pi\Phi/\Phi_0$, where $\Phi$ is the
total magnetic flux enclosed by the path and
$\Phi_0=h/e$ is the magnetic flux quantum. The resulting
current (and conductance) of the quantum ring
is then a periodic function of $\Phi/\Phi_0$.

Recent low-temperature transport experiments~\cite{Heiblum05:abinter,Goldman05:155313,Nissim09:,godfrey:07,Bernd:ABosc.,Marcus09} performed at
two-dimensional (2D) electron systems (2DESs) utilize the
quantum Hall (QH) effect to investigate and control of
the electron dynamics via their AB phase.
An interesting difference between the original AB experiments and
QH interferometers is the fact that in the latter
the electron {\em path} itself may depend on the magnetic field ${\bf B}$.
To describe electron transport in QH interferometers,
the single-particle edge-state approach~\cite{Buettiker86:1761}
is common, but it neglects the dependency of the area enclosed by the
current-carrying channels on the magnetic field~\cite{bernd:07},
as well as on the channel widths. However, as shown explicitly below,
the actual paths can be obtained considering the full many-body
electrostatics, which yields the spatial distribution of
compressible and incompressible strips~\cite{Afif:AB}.

The essential features in the observed AB oscillations in
QH interferometers have been explained using edge-channel
simulations and Coulomb interactions at the classical (Hartree)
level~\cite{Goldman05:155313,Neder06:016804,igor08:ab}.
However, a complete theoretical picture of the observed
phenomenon is still missing~\cite{godfrey:07,goldman:interactions}.
To attain this, it would be
particularly important to (i) describe the full electrostatics
by handling the crystal growth parameters and the ``edge''
definition of the interferometer,
and to (ii) supply this scheme with a dynamical study on
electronic transport in the 2DES.

The objective of this work is to take important steps towards
comprehensive explanation of the AB characteristics
in QH interferometers.
First, we apply the three-dimensional (3D) Poisson equation
and the Thomas-Fermi approximation
for the given heterostructure~\cite{Andreas03:potential},
taking into account the lithographically defined surface patterns.
In this way we obtain the electron and potential distributions
under QH conditions~\cite{Sefa08:prb,Engin:09japon}.
For completeness, we utilize this scheme for the real
experimental geometry resulting from the trench-gating technique.
Second, we determine a model potential describing the current
channels, and use a time-dependent propagation scheme to monitor
transport of a wave packet injected in the channel. We find
distinct AB oscillations, whose characteristics dramatically
depend on the channel widths and their mutual distance.
In particular, we show that there is an optimal way to manipulate
the visibility in realistic AB interferometers.

\section{Device and electrostatics}\label{device}

The current-carrying states in a QH device result
from the Landau-level quantization
followed by level bending at the edges where the Fermi energy crosses
the levels. Thus, the transport takes place through the edge states.
However, there has been substantial debate in the literature whether
the current flows through the compressible or incompressible strips.
Although the ballistic 1D picture~\cite{Halperin82:2185,Buettiker86:1761},
later attributed to compressible strips~\cite{Chklovskii92:4026},
applies well to the integer quantum Hall effect (IQHE),
it requires bulk (localized) states~\cite{Kramer03:172}
to explain the transitions
between the QH plateaus. In contrast, the screening theory
assumes that the current is carried by the scattering-free
incompressible strips only if the widths of these strips (channels)
are wider than the quantum mechanical length scales~\cite{siddiki2004}.

\begin{figure}
\centering
\includegraphics[width=0.7\columnwidth]{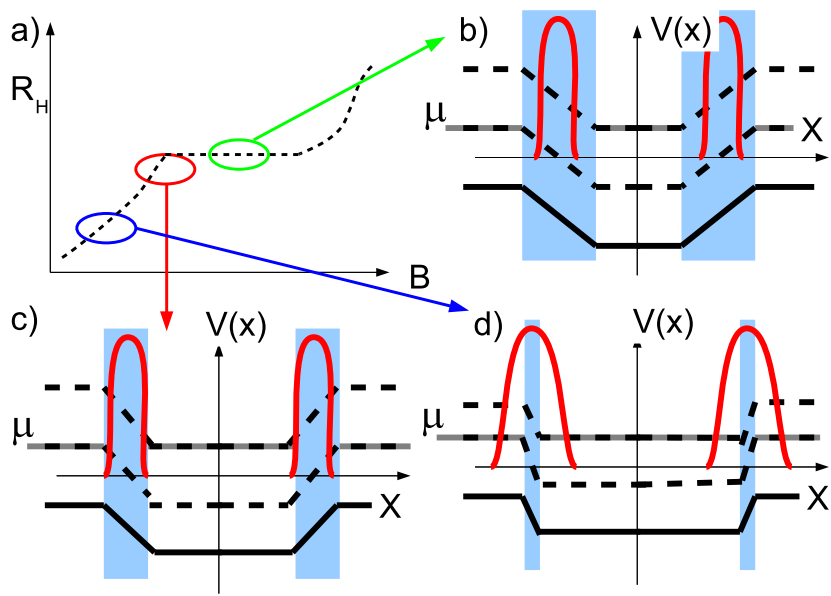}
\caption{(Color online)
(a) Schematic picture of the Hall resistance as a
function of $B$. (b)-(d)
Corresponding potentials (black solid lines),
Landau levels (dashed lines), and wave functions (red solid lines).
Here $\mu$ (gray solid lines) is the Fermi energy at equilibrium.
The ellipses in (a) indicate
the interval in $B$ where the incompressible strips become
well-developed (b), washed out (c), or leaky (d). The actual calculations considering realistic structures can be found in
Refs.~\cite{siddiki2004,SiddikiMarquardt,Sefa08:prb}}\label{fig1}
\end{figure}

A schematic presentation of the Landau levels across a QH bar
is given in Fig.~\ref{fig1}
together with the Hall resistance. There is a steep potential
variation in incompressible strips, where the Fermi energy
falls in between two consecutive Landau levels.
If the incompressible strip is well-developed and larger
than the wave extent,
the quantized Hall effect is observed [Fig.~\ref{fig1}(b)].
However, if the strip width $W_{\rm IS}$ becomes comparable with the wave extent 
(or the magnetic length ($\ell=\sqrt{\hbar/eB}$) for the lowest Landau level)
[Fig.~\ref{fig1}(c)] the strip loses its incompressibility,
and partitioning between channels becomes possible. Thus, it is possible to observe interference. This case is called as the ``leaky" incompressible state, which can be accurately determined by taking into account the collision broadened Landau level width and the extent of the wave function. 
Once the strip becomes even smaller than the wave extent, the classical Hall effect is 
observed~\cite{Ozge:contacts} as shown in Fig.~\ref{fig1}(d). 
Altogether, interference can be observed
at the lower end of the quantized Hall plateau. 

% CUTTED IN THE REVISION PROCESS
%Usually, an interferometer in a high-quality GaAs/AlGaAs
%heterostructure~\cite{Heiblum03:415,Heiblum05:abinter,Goldman05:155313}
%is defined on a ultra-high
%mobility ($\sim 10^6-10^7$ cm$^2$/Vs) wafer by means of etching and/or
%gating~\cite{Heiblum03:415,Roddaro05:156804,Roche07:161309,Litvin07:033315}.
%The current-carrying states~\cite{Halperin82:2185,Buettiker86:1761,Chklovskii92:4026,siddiki2004}
%are utilized as phase-coherent electron ``beams'' to obtain interference
%patterns.  To manipulate the interference pattern, one regulates the
%paths of the current-carrying states by means of side-gate voltages
%at a given $B$ field, {\em or} the gate voltages is fixed and the
%$B$ field is swept.

We calculate of the electron density and the electrostatic potential at the
layer of the 2DES self-consistently~\cite{SiddikiMarquardt} by using the
structural information from Goldman~\cite{Goldman05:155313}.
The dopant density, location
of the interface for the 2DES, and the dielectric constant $\kappa$ (= 12.4 for GaAs/AlGaAs)
are used as the input to calculate the total potential from $V(\vec{r})=V_{\rm conf}(\vec{r})+V_{\rm int}(\vec{r})$,
where the confinement potential $V_{\rm conf}(\vec{r}=(x,y,z))$ is composed of (i)
the potential generated by surface pattern
(corresponding to trench-gating~\cite{Engin:09japon}),
(ii) donors, and (iii) surface charges.
The interaction potential is calculated from the electron density $n_{el}$
by solving the Poisson equation,
\begin{equation}
V_{\rm int}(\vec{r})=\frac{2e^2}{\kappa}\int n_{\rm el}(\vec{r}')K(\vec{r};\vec{r}') d\vec{r}',
\end{equation}
where the kernel $K(\vec{r};\vec{r}')$ takes into account the imposed boundary conditions. The explicit kernels considering different boundary conditions can be found in Refs.~\cite{siddiki2004,Sefa08:prb} and references therein.
The electron density is calculated within the Thomas-Fermi approximation from
\begin{equation}
n_{\rm el}(\vec{r})=\int dE\,D(E;\vec{r})f(V(\vec{r}),E,T,\mu),
\end{equation}
where $D(E;\vec{r})$ is the density of states, and the Fermi occupation function
$f(V(\vec{r}),E,T,\mu)$ depends on the chemical potential $\mu$ and temperature $T$. 
The above equations are solved self-consistently on a uniform 3D grid with
open boundary conditions.

It is important to note the advantages and constraints of the applied approximations.
As a general note, the number of electrons is very large (of the order of $10^3$),
so that semiclassical or density-functional-type approximations become reasonable
in order to obtain a {\em qualitative estimation} for 
the positions and the widths of the current carrying channels. Their
exact properties require formidable numerical approaches that 
are beyond the scope of this work. 
%Esa: NOT NECESSARY
%In any case our approach enables us to 
%describe single particle AB-type interference experiments with reasonable 
%accuracy, in addition, the approximations are viable within the experimental 
%conditions as described below. 
% Esa: I DID NOT UNDERSTAND THIS - THUS COMMENTED OUT
%As a final remark, the leakiness of the 
%incompressible strip can constrain the interference interval to the lower 
%part of the plateau which is not possible within other 
%approaches~\cite{Halperin82:2185,bernd:07}, however, the interference 
%interval may vary in magnetic field within the accuracy of the ratio 
%between the magnetic length and the width of the incompressible strip, 
%i.e. $\ell/W_{\rm IS}$.}

The magnetic field effects are included in our static calculations via the 
approximation of Landau-quantized density of states, although the wave functions 
should also depend on the position according to the total potential. However, 
this approximation can be justified for different regimes of compressibility.
First, in compressible systems the external potential is 
well screened (for instance, similar to the potential at the center of 
Fig.~\ref{fig1}(d), so that the total potential is constant. 
Hence, the Landau wave functions are not modified at all, but the energy 
eigenvalues are shifted. Secondly, in incompressible systems the total potential 
varies linearly, and hence the wave functions preserve the Landau form;
only their center coordinates are shifted slightly (less than the magnetic 
length). At the transition regions the potential may vary with higher-order 
corrections. However, the next (quadratic) correction also preserves the Landau form, 
whereas the center coordinate remains unaffected. In total,
the corrections to the Landau wavefunction are thus negligible up 
to the third order, which justifies the practical validity of our approximation 
in determining the electron density. To compensate for the suppression
of the tunneling effects due to the Thomas-Fermi approximation we
take into account the finite widths of the Landau wave functions as discussed
below.

The exact form of $D(E)$ is essentially determined by the properties of the scattering
 mechanisms in the presence of strong disorder. These constraints are 
lifted by the experimental conditions, so that the number of electrons 
within the interferometer is more than $10^3$ and the mobility of the samples 
is extremely high (typically, $\mu\geq 5\times10^6$ cm$^2$/Vs). Hence, 
in such samples the Landau level broadening can be 
neglected: $\Gamma/(\hbar\omega_c)=\Gamma/(eB/m^{\star})\ll 1$, where 
$\Gamma$ is the level width and $\hbar\omega_c$ is the cyclotron energy.
Otherwise, the level broadening has been proven to have an important influence 
on the (in)compressibility of the system~\cite{Struck}.

We point out that the statistical description of a many-particle system by the
grand-canonical ensemble assumes that the system is in contact with a
reservoir, which can exchange particles with the system and
determines the chemical potential. If the system is
inhomogeneous, one usually considers a constant electrochemical potential
in thermal equilibrium, which, however, is position dependent
in the case of an imposed current, i.e., in a (global) non-equilibrium.

To proceed with the calculation to the desired values of $B$ and $T$,
we impose periodic boundary conditions and replace the constant density
of states in the absence of magnetic fields with a Gaussian-broadened
one~\cite{Sefa08:prb} that takes the
quantization due to the magnetic field into account,
together with disorder. In this procedure,
we first increase $T$ and smear the quantization effects with the
Fermi function. Then, we set the desired value for $B$ and decrease
$T$ stepwise to its target value. In each iteration step a relative
accuracy threshold of $<10^{-6}$ is obtained for the density. This
standard calculation scheme has been previously used to describe similar
systems successfully~\cite{siddiki2004,SiddikiMarquardt,Sefa08:prb,Ozge:contacts}.

\begin{figure}
\centering
\includegraphics[width=0.7\columnwidth]{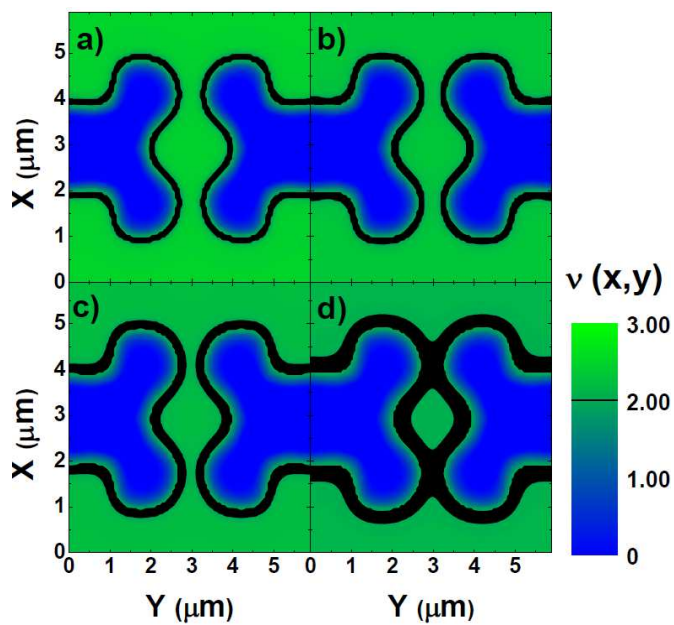}
\caption{Spatial distribution of the incompressible strips (black)
calculated at (a) $B=7.6$ T, (b) $8.0$ T, (c) $8.4$ T, and (d) $8.8$ T
at $T=1$ K. It is expected that only at $B=8.0$ T one can observe
Aharonov-Bohm oscillations,
whereas in the other cases the scattering between
the edge-states is prevented (see text).}
\label{fig2}
\end{figure}

In Fig.~\ref{fig2}
we show the spatial distribution of incompressible strips
at different magnetic fields. The sequence of figures at $B=7.6\ldots 8.8$ T
shows that the distance between the incompressible strips decreases,
and their width increases, as a function of $B$. It can be expected that
at low $B$ when the strips are far apart, the AB oscillations are not present
or they are weak, as the wave packet remains in the incoming channel.
On the other hand, the strong overlap between the strips at
$B=8.8$ T suggests suppression as well -- now due to the complete
transport of the wave packet to the other channel.
Interference would require clean splitting (partitioning) of the wave function 
between the channels. To confirm this phenomenological
suggestion, and to study the effects of the channel distance and width
on the conductance, we focus now on realistic modeling of
current channels along the incompressible strips. This is
followed by a dynamical study on the electron transport
in the device. As the core outcome we can qualitatively describe 
the interference intervals depending on the system parameters such as 
temperature, sample geometry, and heterostructure properties. 
%Such an information is not available in the literature, up to our knowledge.

Before introducing the model for dynamics, a brief discussion
about the direction of the current along the incompressible
strips is in order. It is well known that the current-carrying
states in a QH device result from the Landau-level quantization
followed by level bending at the edges where the Fermi energy crosses
the levels. Thus, the \emph{equilibrium} transport takes place
through the chiral edge states, i.e., with different current
directions at opposite edges, and the net current is zero.
In transport experiments (including interference experiments),
however, an external current is imposed and a Hall (electrochemical)
potential difference -- with the {\em same slope} -- develops between two opposing
edges~\cite{Guven03:115327}. This potential drop at the
incompressible strips, that confines the external current in those
regions, has been shown in several
experiments~\cite{Ahlswede01:562,Ahlswede02:165,Dahlem:10}.
In the following we utilize this picture in the transport
calculations, so that the electrons flow to the same direction along
the opposite channels.

\section{Electron transport}

%Next we examine the time-propagation of an electron injected
%in the transport channels corresponding to the incompressible
%strips described above.

\begin{figure}
\centering
\includegraphics[width=0.8\columnwidth]{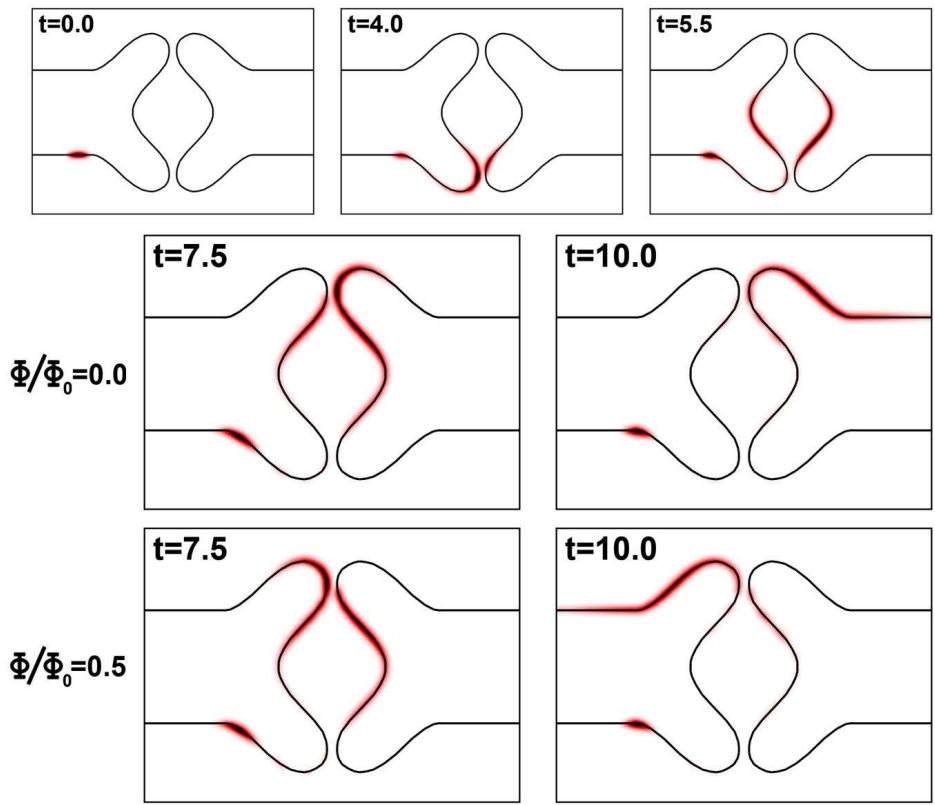}
\caption{Snapshots of the electron density in a model for the Aharonov-Bohm
interferometer. The two lowermost rows correspond to zero flux and half
a flux quantum, respectively.}
\label{fig3}
\end{figure}

The channels are modeled by a 2D potential profile consisting
of two curved pipes following the shape of the incompressible
strips in Fig.~\ref{fig2}. The potential {\em minima} of the
channels are shown as solid lines in Fig.~\ref{fig3}. The
distance between the left and right parts is varied such that
at the two encountering points (bottom and up) the distance between the potential
minima is $d=0\ldots 1$ (in atomic units). It is important
to note that the channels are genuinely 2D according to the
real device, and their cross sections have a Gaussian form
$V_{\rm cross}\approx -V_0\,e^{-s^2/c^2}$, where $s$ is the coordinate
perpendicular to the channel, $V_0=20$ is the channel depth, and $c=0.2$ is
the width parameter which is varied. The choice of a Gaussian profile
is justified by considering the magnetic (parabolic) confinement, which is
close to a Gaussian form at the bottom of the channel. On the other hand,
in the upper part of the channel the selected profile allows ``leaking''
of the electron flow according to the experiment (see above).

As an initial condition we set a single-electron wave packet in the lower-left
corner of the device and, in order to emulate a source-drain voltage, we
accelerate the wave packet using a linear potential with slope $V_1=-0.2 V_0$
along the channel, which leads to an initial velocity of $\sim 3.3$ a.u.
We monitor the density and the current density during the time propagation
until we find back-scattering form the upper left and right corners of our
{\em finite} simulation box. In this way, we guarantee the correct
direction of the current along the incompressible strips (see the discussion
above).
The conductance as a function of the magnetic
field, the channel distance, and the channel width is
estimated by calculating the
probability $N_r$ to find the electron in the upper-right corner within the
restricted simulation time (see Ref.~\cite{Esa:ab} for details).
We made the time propagation by using the fourth order Taylor expansion of time evolution operator to the
single-electron wave function in a 2D real-space grid with the {\tt octopus}
code package~\cite{octopus}.

We emphasize that in contrast to the real interferometer (see Fig.~\ref{fig2}),
where both the channel width and their mutual distance (i.e., the spatial
distribution of the incompressible strips)
is determined by the magnetic field {\em itself},
we let in our model the magnetic field affect only the flux but
not the distribution of the channels. However, for each set of
calculations, respectively, the width and distance are changed
through the model parameters given just above.

In our transport simulations we monitor electron dynamics as a function
of a magnetic flux {\em added} to the background magnetic field (the
latter generating the incompressible strips).
In Fig.~\ref{fig3} we show snapshots of
the electron density at different
times. Here we apply relative fluxes $\Phi/\Phi_0=0$ and 0.5,
respectively. In the
beginning (see the uppermost row) differences in density
between these two cases are not visible, so only the other case is plotted.
Later on, however, we find a drastic difference between the two cases:
whereas zero flux leads to transport to the right (second row), a flux
corresponding to half a flux quantum yields strong current to the left (third row).
Thus, by exploiting the AB effect, our device is almost completely tunable with
respect to the current direction at the ouput (left or right).

\begin{figure}
\centering
\includegraphics[width=0.6\columnwidth]{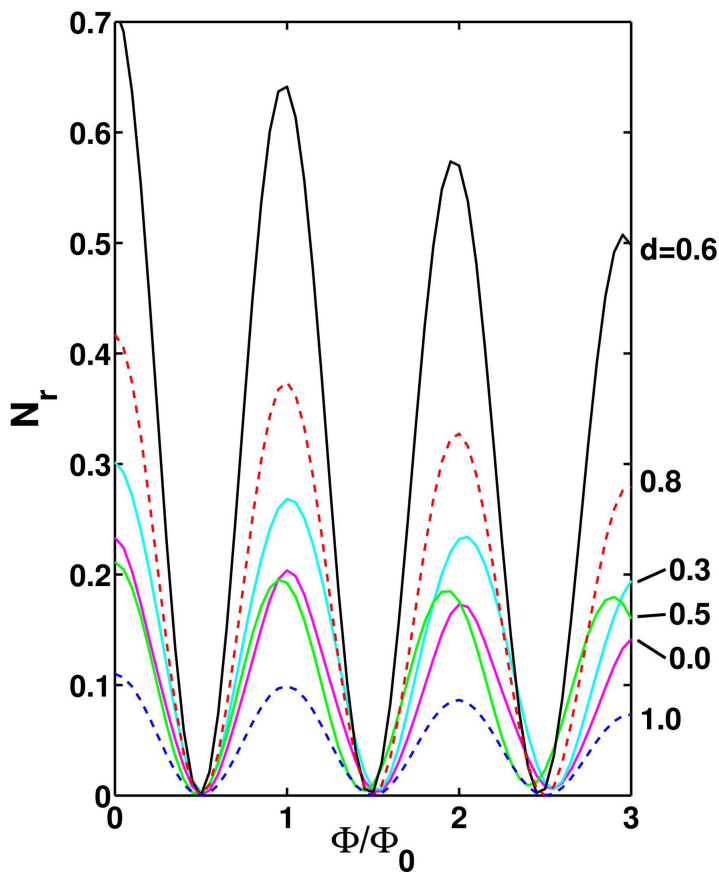}
\caption{Estimated conductance $N_r$ as a function of the magnetic flux
in the model interferometer with different distances between the channels.}
\label{fig4}
\end{figure}

To analyze the transport quantization in more detail, we plot in
Fig.~\ref{fig4}
the conductance -- corresponding here to the electron density
transferred to the top of the right channel -- as a function of
the magnetic flux,
and for different distances between the channels. We find clear and smooth
AB oscillations having exactly the expected periodicity. It is noteworthy
that in all cases the interchannel transport is completely blocked at
$\Phi/\Phi_0=k/2$, where $k$ is an odd integer.

\begin{figure}
\centering
\includegraphics[width=0.6\columnwidth]{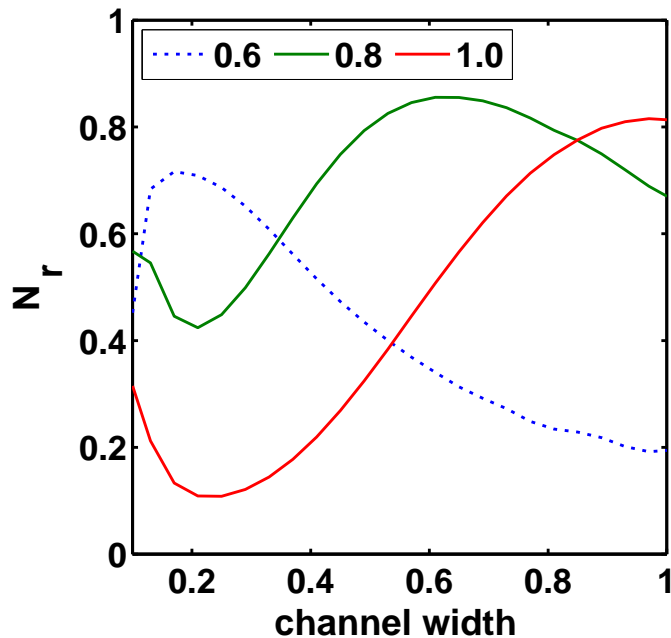}
\caption{Estimated conductance $N_r$ at zero magnetic field
as a function of the channel width parameter $c$ for
different distances ($d$) between the channels.}
\label{fig5}
\end{figure}

An interesting feature in Fig.~\ref{fig4} is the fact that the channel
distance has an {\em optimal} distance $d\sim 0.6$ when the AB oscillation amplitude
is maximized. At smaller (solid lines) and larger (dashed lines) values of $d$
the amplitude is decreased. We find that the optimal distance corresponds to
a case when approximately a half of the density is transferred to the right
channel at the first (lower) intersection, i.e., the partitioning is equal.
Therefore, at the second (upper) intersection we find clear AB interference
due to the phase difference of the {\em optimally partitioned} electrons that
enclose the given flux. Our results agree with the behavior in the AB oscillation
strength observed in real AB interferometers~\cite{Roche07:161309,Litvin07:033315}.
We note, however, that as demonstrated in Fig.~\ref{fig5}, the optimal distace depends
on the channel width $c$ in a nontrivial fashion -- apart from the large-$c$ limit showing
clear suppression of the amplitude. However, it seems that half-half partitioning
of the density distribution at the first (bottom) point of encounter always
leads to a relatively high amplitude of the AB oscillation. A complete
{\em quantitative} comparison with experiments would require an accurate determination
of the system geometry as a function of the magnetic field.

\section{Summary}

To summarize, we have performed static and dynamical simulations
on Aharanov-Bohm interferometers starting from real device parameters.
First, the electron density and the spatial distribution of the incompressible
strips have been obtained self-consistently within the Thomas-Fermi approximation.
These calculations already suggest that interference can take place only if the
incompressible strips become leaky, namely not strictly compressible or 
incompressible, and come close to each other, so that
partitioning of the electron current can take place. These phenomenological
considerations have then been qualitatively confirmed in the second part of the study, where
we time-propagate an electronic wave function injected in the channel with
tunable parameters. We observe distinct Aharonov-Bohm oscillations, whose
amplitude strongly depends on both the mutual distance between the transport
channels and their width. In particular, there is an optimal distance yielding
maximum oscillation amplitudes. At magnetic fields corresponding to
half-integer flux quanta the suppression of interchannel transport is complete.
Taken together, we are able to provide a proof-of-concept for the determination
of interference patterns in realistic Aharonov-Bohm interferometers in the 
quantum Hall regime.

\ack
This work was supported by the Magnus Ehrnrooth Foundation,
the Academy of Finland, TUBITAK:109T083, IU-BAP:6970, and the
Wihuri Foundation. CSC Scientific Computing Ltd.
is acknowledged for computational resources.
A.S. would like to thank to M. Heiblum, N. Offek, and L. Litvin for
experimental discussions. We are also grateful to V.J.
Goldman for providing us the sample structure and device pattern.

\end{document}